\documentclass[twocolumn,prd,nofootinbib]{revtex4}
\usepackage{makeidx}
\usepackage{amsmath}
\begin{document}

\title{Asymmetric brane-worlds with induced gravity}

\author{L\'{a}szl\'{o} \'{A}. Gergely$^{1,2}$ and Roy Maartens$^{1}$}

\affiliation{$^{1}$Institute of Cosmology and Gravitation,
University of Portsmouth, Portsmouth PO1 2EG, UK }

\affiliation{$^{2}$Departments of Theoretical and Experimental
Physics, University of Szeged, Szeged 6720, D\'{o}m t\'{e}r 9,
Hungary}

\begin{abstract}

The Randall-Sundrum scenario, with a 1+3-dimensional brane in a
5-dimensional bulk spacetime, can be generalized in various ways.
We consider the case where the $Z_2$-symmetry at the brane is
relaxed, and in addition the gravitational action is generalized
to include an induced gravity term on the brane. We derive the
complete set of equations governing the gravitational dynamics for
a general brane and bulk, and identify how the asymmetry and the
induced gravity act as effective source terms in the projected
field equations on the brane. For a Friedmann brane in an anti de
Sitter bulk, the solution of the Friedmann equation is given by
the solution of a quartic equation. We find the perturbative
solutions for small asymmetry, which has an effect at late times.

\end{abstract}

\date{\today }

\maketitle

\section{Introduction}

Developments in particle physics suggest that our observable world
may be a 1+3-dimensional ``brane" surface embedded in a
higher-dimensional ``bulk" spacetime, with matter fields confined
on the brane while gravity propagates in the bulk. An important
family of such brane-world models is provided by the
Randall-Sundrum (RS) scenario, with a single warped and
non-compact extra dimension~\cite{rev}. The effective Einstein
equations on the brane in the general case were derived via the
covariant Shiromizu-Maeda-Sasaki approach~\cite{SMS}, together
with equations governing the bulk degrees of freedom. A
generalization of these equations for the case when the $Z_{2}$
-symmetry of the embedding is lifted and the bulk contains matter,
was recently presented in~\cite{Decomp}. Due to the asymmetry new
terms appear in the effective Einstein equations, including a
varying cosmological ``constant''.

A correction arising from the quantum interaction of bulk
gravitons and brane matter can be introduced into RS models by
adding the so-called induced gravity contribution to the
gravitational action:
\begin{eqnarray}
S &=&\frac{1}{2\widetilde{\kappa }^{2}}\int
d^{5}x\sqrt{-\widetilde{g}}\left[
\widetilde{R}-2\widetilde{\Lambda }\right]   \notag \\
&&~{}+\int_{\mbox{brane}}d^{4}x\sqrt{-g}\left[ -\lambda
+\frac{\gamma }{ 2\kappa ^{2}}\,R\right] .  \label{action}
\end{eqnarray}
Tildes indicate the 5-dimensional counterparts of standard
4-dimensional quantities, $\lambda $ is the brane tension and
$\gamma $ is the dimensionless induced-gravity parameter, with
$\gamma =0$ giving the standard RS model. The brane gravitational
constant is given by~\cite{SMS}
\begin{equation}
\kappa ^{2}=\frac{\widetilde{\kappa }^{4}\lambda }{6}\,.
\label{kappa}
\end{equation}
The effective Einstein equations (with $Z_{2}$-symmetry) were
derived for the general case in~\cite{d1,MMT}, generalizing the
earlier special cases, such as the Dvali-Gabadadze-Porrati
models~\cite{DGP,d2}, in which $\gamma =1$ and Eq.~(\ref{kappa})
does not hold, since $\lambda =0=\widetilde{\Lambda }$. The
induced gravity models contain two branches, labelled $\epsilon
=\pm 1$, with very different properties. The $\epsilon =-1$ branch
has an RS limit, including the case where induced gravity is a
sub-dominant correction to the RS model~\cite{ig,KLM,Shtanov}.

In the present paper we lift the $Z_{2}$-symmetry of the
embedding, and derive the complete system of equations, including
the effective Einstein equations on the brane. Some earlier
results for this scenario in the particular case of Friedmann
branes embedded in a vacuum bulk were presented
in~\cite{KLM,Shtanov}.

In Section 2, by use of the Lanczos-Sen-Darmois-Israel junction
conditions~\cite{Lanczos,Sen,Darmois,Israel}, we derive the
generalized effective Einstein equations in a form consistent with
previous works. We show that, remarkably, the asymmetric
contributions add \textit{linearly} to the induced gravity
corrections.

In Section 3 we apply the formalism to the case of Friedmann
branes. We derive the generalized Friedmann and Raychaudhuri
equations together with the constraints emerging from the Codazzi
equation and other equations of the asymmetric case. Among the
asymmetric features, we find a variable 4-dimensional cosmological
``constant" $\Lambda $ and contributions of the embedding to the
dark radiation $U$. As in the $Z_2$-symmetric induced gravity
models, the non-linearity of the Friedmann equation in $H^2$ leads
to two branches, $\epsilon =\pm 1$. The asymmetric RS model can be
recovered in the $\epsilon =-1$ case, as the $\gamma\rightarrow 0$
limit.

Following the method of~\cite{Decomp}, in Section 4 we present an
independent derivation of the Friedmann equation in the simplest
case when the bulk is 5-dimensional anti de Sitter space-time
(AdS), with \textit{\ different} cosmological constants on the two
sides of the brane. The comparison of the two forms of the
Friedmann equation yields a quartic equation in $H^2$, as
in~\cite{KLM,Shtanov}. We find a perturbative solution of the
quartic equation for small asymmetry. Then, by use of the twice
contracted Bianchi identity we identify both unknown {functions}
$\Lambda $ and $U$. Finally we analyze the high- and low-energy
limits of the Friedmann equation.

\section{Effective Einstein equations}

\subsection{The general projection formalism}

In this subsection we summarize the relevant results
of~\cite{Decomp}.

The first and second fundamental forms of the brane are
\begin{eqnarray}
g_{ab} &=&\widetilde{g}_{ab}-n_{a}n_{b}\,,  \label{tildeg0} \\
K_{ab} &=&\nabla _{a}n_{b}={\frac{1}{2}}\mathcal{L}_{\mathbf{n}}
g_{ab}\,, \label{KLie}
\end{eqnarray}
where $n^a$ is the unit normal to the brane, with acceleration $
\alpha ^{a}=n^{b}\widetilde{\nabla }_{b}n^{a}=g_{b}^{a}\alpha
^{b}$. Following~\cite {Decomp} we introduce the tensors
\begin{align}
J_{ab}& =K_{ac}K_{b}^{c}-\mathcal{L}_{\mathbf{n}}K_{ab}+\nabla
_{b}\alpha
_{a}-\alpha _{b}\alpha _{a}\,,  \label{E} \\
F_{ab}& =KK_{ab}-K_{ac}K_{b}^{c}\,.  \label{F}
\end{align}
As shown in~\cite{Decomp}, the 5-dimensional Einstein equations
$\widetilde{G }_{ab}=-\widetilde{\Lambda}\widetilde{g}_{ab}$ are
equivalent to the system comprised of the effective 4-dimensional
Einstein equations, the Codazzi equation and the twice contracted
Gauss equation:
\begin{eqnarray}
G_{ab}&=& \frac{2}{3}\widetilde{G}_{\{ab\}} + {\frac{1}{2}}
n^{c}n^{d}
\widetilde{G}_{cd}\,g_{ab}  \notag \\
&&~{}+F_{ab}-{\frac{1}{2}}F{g_{ab}} -\mathcal{E}_{ab}\,,
\label{modE} \\
\nabla _{b}K_{a}^{b}-\nabla _{a}K &=&
g_{a}^{b}{}n^{c}{}\widetilde{G}_{bc}\,,
\label{Codazzi} \\
R&=& \widetilde{R}+F-2J\,.  \label{Gauss2}
\end{eqnarray}
Here
\begin{equation}
Q_{\{ab\}}\equiv \left(g_a^cg_b^d-{\frac{1}{4}}g_{ab}g^{cd}\right)
Q_{cd}\,,
\end{equation}
denotes the projected trace-free part, and the ``electric''\ part
of the bulk Weyl tensor is
\begin{equation}
\mathcal{E}_{ab} \equiv \widetilde{C}_{acbd}n^cn^d
=J_{\{ab\}}-{\frac{1}{3}} \widetilde{G}_{\{ab\}}\,. \label{eps1}
\end{equation}

The brane divides the bulk into two distinct regions. The jump and
average of a quantity across the brane are $\Delta Q=Q^{+}-Q^{-}$
and $\overline{Q}={\frac{1}{2}}(Q^{+}+Q^{-})$. The junction
conditions impose continuity of the first fundamental form,
$\Delta g_{ab}=0$, and the Lanczos equation~\cite {Lanczos}
\begin{equation}
\Delta K_{ab}=-\widetilde{\kappa }^{2}\left( \tau
_{ab}-{\frac{\tau }{3}}g_{ab}\right) \,,  \label{Lanczos}
\end{equation}
which relates the jump of the second fundamental form to the total
brane energy-momentum. (These conditions were first formulated in
a coordinate-independent manner in~\cite{Israel}.)
Then~\cite{Decomp}:
\begin{eqnarray}
\Delta F_{ab} &=&-\widetilde{\kappa }^{2}\Bigl[\overline{K}\left(
\tau
_{ab}-\frac{\tau }{3}g_{ab}\right)   \notag \\
&&~{}+\frac{\tau }{3}\overline{K}_{ab}-2\overline{K}_{c(a}\tau
_{b)}^{c}\Bigr]\,, \\
\overline{F}_{ab}
&=&\overline{K}_{ab}\overline{K}-\overline{K}_{ac}
\overline{K}_{b}^{c}+\delta F_{ab}\,, \\
\delta F_{ab} &\equiv &-\frac{\widetilde{\kappa }^{4}}{4}\left(
\tau _{ac}\tau _{b}^{c}-\frac{\tau }{3}\tau _{ab}\right) \,.
\end{eqnarray}
The sums of the effective Einstein equation, Codacci equation and
twice contracted Gauss equation, taken on the two sides of the
brane, lead to
\begin{eqnarray}
G_{ab} &=&-{\frac{1}{2}}\overline{\tilde{\Lambda
}}g_{ab}+\widetilde{ \kappa }^{2}\left(
\overline{F}_{ab}-\frac{\overline{F}}{2}g_{ab}-\overline{
\mathcal{E}}_{ab}\right) \,,  \label{modEp} \\
\nabla _{b}\overline{K}_{a}^{b} &=&\nabla _{a}\overline{K}\,,
\label{vp}
\\
R &=&\frac{10}{3}\overline{\tilde{\Lambda}}+\overline{F}
-2\overline{J}\,. \label{G2p}
\end{eqnarray}
The differences of the same equations give:
\begin{eqnarray}
\Delta \mathcal{E}_{ab} &=&\Delta F_{\{ab\}}\,,  \label{trlessm} \\
\widetilde{\kappa }^{2}\overline{K}_{ab}\tau ^{ab} &=&-\Delta
\widetilde{
\Lambda }\,,  \label{trm} \\
\nabla _{b}\tau _{a}^{b} &=&0\,,  \label{vm} \\
\widetilde{\kappa }^{2}\overline{K}_{ab}\tau ^{ab} &=&\Delta
J-\frac{5}{3} \Delta \widetilde{\Lambda }\,.  \label{G2m}
\end{eqnarray}
We have further decomposed the difference of the effective
Einstein equations into tracefree and trace parts, in
Eqs.~(\ref{trlessm}) and (\ref {trm}).

\subsection{Induced gravity equations}

Up to this point, all results are unchanged compared to the
asymmetric RS model, presented in~\cite{Decomp} (here we have
omitted the arbitrary bulk energy-momentum tensor
in~\cite{Decomp}). This is simply because induced gravity does not
modify bulk physics, as seen in Eq.~(\ref{action}). However,
Eq.~(\ref{action}) also shows that the induced gravity scenario
adds a new term to the brane total energy-momentum~\cite{MMT}:
\begin{equation}
\tau _{ab}=-\lambda g_{ab}+T_{ab}-{\frac{\gamma
}{\kappa^2}}G_{ab}\,, \label{induced}
\end{equation}
where $T_{ab}$ represents ordinary matter on the brane. Then it
follows that
\begin{align}
\delta F_{ab}-{\frac{\delta F}{2}}g_{ab}&=\kappa^{2}\left(
-\frac{\lambda }{2
}g_{ab}+T_{ab}\right)- {\gamma} G_{ab}  \notag \\
&~~{}+{\frac{6}{\kappa^2\lambda }}\mathcal{S}[\kappa^2T-\gamma
G]_{ab}\,,
\end{align}
where the tensor functional $\mathcal{S}[Q]$ is defined
by~\cite{MMT}
\begin{eqnarray}
\mathcal{S}[Q]_{ab}&=& \frac{1}{4}\Biggl[-Q_{ac} Q_{b}^{c}
+\frac{Q}{3}Q_{ab}
\notag \\
&&~{} -{\frac{1}{2}}g_{ab}\left(
-Q_{cd}Q^{cd}+\frac{Q^{2}}{3}\right) \Biggr] \,.  \label{S}
\end{eqnarray}
The tensor $\mathcal{S}[T]_{ab}$ that is quadratic in $T_{ab}$,
was introduced for the $Z_2$-symmetric case in~\cite{SMS}.

We define the brane cosmological ``constant''~\cite{Decomp}:
\begin{equation}
\Lambda =\Lambda _{0}-\frac{\overline{L}}{4}\,,  \label{Lambda}
\end{equation}
where $\Lambda _{0}$ is a constant,
\begin{equation}
\Lambda _{0}=\frac{\kappa ^{2}\lambda
}{2}+\frac{\overline{\tilde{\Lambda}}}{ 2}\,,
\end{equation}
while $\overline{L}$ is a variable determined by the asymmetric
embedding, being the trace of the following tensor:
\begin{equation}
\overline{L}_{ab}\equiv
\overline{K}\overline{K}_{ab}-\overline{K}_{ac}
\overline{K}_{b}^{c}-\frac{g_{ab}}{2}\left(
\overline{K}^{2}-\overline{K} _{cd}\overline{K}^{cd}\right) \,.
\end{equation}
Then the {effective Einstein equations} are
\begin{align}
\left( 1+\gamma \right) G_{ab}& =-\Lambda g_{ab}+\kappa
^{2}T_{ab}-\overline{
\mathcal{E}}_{ab}  \notag \\
& ~{}+{\frac{6}{\kappa ^{2}\lambda }}\mathcal{S}[\kappa
^{2}T-\gamma G]_{ab}+ \overline{L}_{\{ab\}}\,.  \label{modEgen}
\end{align}
The asymmetric contributions~\cite{Decomp} in $\Lambda $ and in
the source term $\overline{L}_{\{ab\}}$ add up \textit{linearly}
with the induced gravity contributions, derived first
in~\cite{MMT}. Note that, as defined here, the cosmological
``constant''\ is a \textit{function} of the embedding of the
brane.

Both the Codazzi equation~(\ref{vp}) and Eq.~(\ref{G2p}), giving
$\overline{J }$, are insensitive to brane physics, such as the
choice of the matter energy-momentum tensor and the presence of
the induced gravity correction. However the following equations do
depend on these features. Using Eq.~(\ref {induced}) in
Eq.~(\ref{trlessm}) and combining Eqs. (\ref{trm}) and (\ref
{G2m}), we obtain $\Delta \mathcal{E}_{ab}$ and $\Delta J$:
\begin{eqnarray}
{\frac{\kappa ^{2}}{\widetilde{\kappa }^{2}}}\Delta
\mathcal{E}_{ab} &=&- \overline{K}\left[ {\kappa
}^{2}T_{\{ab\}}-\gamma G_{\{ab\}}\right] \notag
\\
&&~{}-{\frac{1}{3}}\overline{K}_{\{ab\}}{\left[ {\kappa
}^{2}(2\lambda
+T)-\gamma G\right] }  \notag \\
&&~{}+2\overline{K}_{c\{a}\left[ {\kappa }^{2}T_{b\}}^{c}-\gamma
G_{b\}}^{c}
\right] \,,  \label{dep} \\
\Delta J &=&{\frac{2}{3}}\Delta \widetilde{\Lambda }\,.
\label{de}
\end{eqnarray}
These equations, together with Eq.~(\ref{G2p}), characterize the
off-brane evolution of $K_{ab}$, and we will not deal with them
further here.

In the asymmetric case there are also the additional constraints,
Eqs.~(\ref {trm}) and (\ref{vm}), to be imposed on the matter
fields and geometrical properties of the brane:
\begin{eqnarray}
\nabla _{b}T_{a}^{b} &=&0\,,  \label{constr2} \\
-\lambda \overline{K}+\overline{K}^{ab}\left(
T_{ab}-{\frac{\gamma}{\kappa^2} } G_{ab}\right)
&=&-{\frac{1}{\widetilde{\kappa}^2}} \Delta\widetilde { \Lambda}
\,.  \label{constr1}
\end{eqnarray}
From the twice-contracted Bianchi identity in 4 dimensions, using
Eq.~(\ref {constr2}), we find
\begin{gather}
\nabla ^{a}\left(
\overline{\mathcal{E}}_{ab}-\overline{L}_{\{ab\}}\right) =-
{\frac{6}{\kappa^2\lambda }} \nabla^a \mathcal{S}[\kappa^2T-\gamma
G]_{ab} \,.  \label{Bianchi}
\end{gather}
The cosmological consequences of this equation will be exploited
in the next section.

\section{Friedmann brane}

For a Friedmann brane,
\begin{eqnarray}
g_{ab} &=&-u_{a}u_{b}+a^{2}h_{ab}\,,  \label{fmetric} \\
T_{ab} &=&\rho u_{a}u_{b}+pa^{2}h_{ab}\,,  \label{fenmom} \\
G_{ab} &=&\frac{3(\dot{a}^{2}+k)}{a^{2}}u_{a}u_{b}-\left(
2a\ddot{a}+\dot{a} ^{2}+k\right) h_{ab}\,,  \label{fG}
\end{eqnarray}
where $u^{a}$ is the geodesic 4-velocity, $a$ is the scale factor
and $h_{ab} $ is the 3-metric with constant curvature (and
curvature index $k=0,\pm 1$) of the maximally symmetric spacial
slices, on which the energy density $\rho $ and pressure $p$ are
constant. Then Eq.~(\ref{S}) implies
\begin{equation}
\mathcal{S}[\kappa ^{2}T-\gamma
G]_{ab}=Vu_{a}u_{b}+Wa^{2}h_{ab}\,, \label{fS}
\end{equation}
where
\begin{eqnarray}
V &=&\left[ {\frac{\kappa ^{2}\rho
}{\sqrt{12}}}-{\frac{\sqrt{3}\gamma (\dot{
a}^{2}+k)}{2a^{2}}}\right] ^{2}, \\
W &=&{\frac{\kappa ^{4}\rho (\rho +2p)}{12}}-{\frac{\gamma \kappa
^{2}[(2\rho +3p)(\dot{a}^{2}+k)-2\rho a\ddot{a}]}{6a^{2}}}  \notag \\
&&~{}+{\frac{\gamma
^{2}(\dot{a}^{2}+k)(\dot{a}^{2}+k-4a\ddot{a})}{4a^{4}}} \,.
\end{eqnarray}

We can introduce an effective ``non-local'' energy density $U$,
which imprints bulk effects on the brane (generalizing the
$Z_{2}$-symmetric procedure~\cite{Maartens2}), via
\begin{equation}
-\overline{\mathcal{E}}_{ab}+\overline{L}_{\{ab\}}=\kappa
^{2}U\left( u_{a}u_{b}+\frac{a^{2}}{3}h_{ab}\right) \,.  \label{U}
\end{equation}
In the $Z_{2}$-symmetric case, $U$ arises from the Coulomb field
of a bulk black hole, and is called ``dark radiation''; for an AdS
bulk, $U=0$. Asymmetry allows for a non-zero $U$ in an AdS bulk.
Then the effective Einstein equation~(\ref{modEgen}) gives the
generalized Friedmann equation,
\begin{gather}
\left[ 1+\gamma \left( 1+\frac{\rho }{\lambda }\right) \right]
\frac{(\dot{a} ^{2}+k)}{a^{2}}=\frac{3\gamma ^{2}}{2\lambda \kappa
^{2}}\left( \frac{\dot{a}
^{2}+k}{a^{2}}\right)^{2}  \notag \\
{}+\frac{1}{3}\left[ \Lambda +\kappa ^{2}\rho \left( 1+\frac{\rho
}{2\lambda }\right) +\kappa ^{2}U\right] \,,  \label{Fried}
\end{gather}
and the generalized Raychaudhuri equation,
\begin{align}
& \left[ 1+\gamma \left( 1+\frac{\rho }{\lambda }\right) \right]
\frac{ \ddot{a}}{a}-{3\gamma }\left( \frac{\rho +p}{2\lambda
}\right) \frac{(\dot{a}
^{2}+k)}{a^{2}}=  \notag \\
& ~{}\frac{1}{6}\Biggl\{2\Lambda -\kappa ^{2}\Bigl[3p\left(
1+\frac{\rho }{ \lambda }\right) +\rho \left( 1+2\frac{\rho
}{\lambda }\right) \Bigr
]-2\kappa ^{2}U\Biggr\}  \notag \\
& ~{}+\frac{3\gamma ^{2}}{2\lambda \kappa ^{2}}\left(
\frac{\dot{a}^{2}+k}{ a^{2}}\right) \left(
2\frac{\ddot{a}}{a}-\frac{\dot{a}^{2}+k}{a^{2}}\right) \,.
\label{Raych}
\end{align}
These have the same form as Eqs.~(3.20,21) of~\cite{MMT}. All
asymmetric features are absorbed in $\Lambda $ and $U$, which
{are} different from those of~\cite{MMT}. The conservation
equation~(\ref{constr2}) has the usual form,
\begin{equation}
\dot{\rho}+3\frac{\dot{a}}{a}\left( \rho +p\right) =0\,,
\label{rhodot}
\end{equation}
and the other constraint~(\ref{constr1}) with no symmetric
counterpart gives
\begin{align}
& {\frac{1}{\widetilde{\kappa }^{2}}}\Delta \widetilde{\Lambda
}=\left[ \lambda -p-{\frac{\gamma }{\kappa ^{2}}}\left(
2\frac{\ddot{a}}{a}+\frac{
\dot{a}^{2}+k}{a^{2}}\right) \right] \overline{K}  \notag \\
& ~{}-\left[ \rho +p+2{\frac{\gamma }{\kappa ^{2}}}\left(
\frac{\ddot{a}}{a}- \frac{\dot{a}^{2}+k}{a^{2}}\right) \right]
u_{a}u_{b}\overline{K}^{ab}\,. \label{fc1}
\end{align}
In the RS model ($\gamma \rightarrow 0$ limit) this is just a
constraint on the embedding and the perfect fluid.
Induced gravity introduces\ $\dot{a}$ and
$\ddot{a}$ contributions, transforming Eq.~(\ref{fc1}) into a
dynamical equation, to be added to the system of the
energy-balance equation~(\ref{rhodot}) and the generalized
Friedmann and Raychaudhuri equations~(\ref{Fried}) and
(\ref{Raych}). We will discuss the significance of this equation
below.

The ``non-local" conservation equation~(\ref{Bianchi}) reduces to
\begin{equation}
\dot U+ 4{\frac{\dot a}{a}}\,U={\frac{1}{4\kappa^2}}\dot{\overline
L}\,, \label{U0}
\end{equation}
which shows how asymmetry can be a source for $U$ even in the
absence of a bulk black hole. For a $Z_{2}$-symmetric embedding,
we have $\overline L=0$ and $U\propto a^{-4}$. An independent way
to derive Eq.~(\ref{U0}) is to take the time derivative of the
generalized Friedmann equation, then eliminate $\dot{\rho}$ by
Eq.~(\ref{rhodot}) and eliminate the derivatives of $a$ from the
terms not containing $\gamma$ by Eqs.~(\ref{Fried}) and
(\ref{Raych}).

The Lanczos equation becomes
\begin{gather}
\Delta K_{ab}=-{\frac{\widetilde{\kappa }^{2} }{3}}\Biggl\{\left[
2\rho +3p- {\lambda }+{\frac{3\gamma }{\kappa^2}} \left(
2\frac{\ddot{a}}{a}-\frac{\dot{
a}^{2}+k}{ a^{2}}\right) \right] u_{a}u_{b}  \notag \\
{}+\left[ {\rho +\lambda }-{\frac{3\gamma }{\kappa^2}}\left(
\frac{\dot{a} ^{2}+k}{a^{2}} \right) \right] a^{2}h_{ab}\Biggr\}.
\label{fLanczos}
\end{gather}

Up to this point the discussion has followed~\cite{Decomp}, the
results of which can be recovered in the $\gamma\rightarrow 0$
limit. Since the Friedmann equation~(\ref{Fried}) is quadratic in
$\left( \dot{a} ^{2}+k\right) $, we can take its square root,
following~\cite{MMT}, and obtain
\begin{equation}
H^2+\frac{k}{a^{2}}= \frac{\kappa^2}{3\gamma}\left[ \rho +
{\frac{(1+\gamma) }{\gamma}}\,\lambda \left( 1+\epsilon
\mathcal{A}\right) \right] , \label{Fr}
\end{equation}
where
\begin{equation}
\mathcal{A}^{2}= 1+\frac{2\gamma}{(1+\gamma)^2\lambda}\left[ \rho
-{\frac{ \gamma}{\kappa^2}}\left( \Lambda +\kappa ^{2}U\right)
\right] , \label{A2}
\end{equation}
and $\epsilon =\pm 1$ distinguishes the two branches. Although the
above equations have the same form as those in~\cite{MMT}, the
function $\mathcal{ A=A}\left( \rho ,U,\Lambda\right) $ depends on
the asymmetric character of the embedding. Employing
Eq.~(\ref{Fr}), the Raychaudhuri equation~(\ref {Raych}) can also
be simplified,
\begin{align}
&-6\epsilon(1+\gamma)\mathcal{A}\,\frac{\ddot{a}}{a}=
{\frac{\kappa^2}{\gamma
}}(\rho+3p)+2(\Lambda-\kappa^2U)  \notag \\
&{} -{\kappa
^{2}{\frac{(1+\gamma)^2}{\gamma^2}}\,\lambda}(1+\epsilon
\mathcal{A})^2 + \epsilon{\kappa
^{2}}{\frac{(1+\gamma)}{\gamma}}\,\left( \rho +3p\right)
\mathcal{A}\,.  \label{Ray}
\end{align}

It is not immediate to obtain the RS limit from the above
equations, since we need to go to second order in $\gamma$. In the
limit of small induced gravity,
\begin{eqnarray}
(1+\gamma)\mathcal{A} &\rightarrow &1+{\gamma}\left(1+ {\frac{\rho
}{\lambda}
}\right)  \notag \\
&&{}-{\frac{\gamma^2}{\lambda}}\left[ \rho \left( 1+\frac{\rho
}{2\lambda } \right) +{\frac{\Lambda}{\kappa^2}} +U\right],
\label{limit}
\end{eqnarray}
and to leading order we recover the Friedmann and Raychaudhuri
equations of the asymmetric RS model in the $\epsilon =-1$ branch.
The other branch does not allow this limit.

We can now eliminate the derivatives of $a$ from Eq.~(\ref{fc1}),
to obtain an algebraic relation among $U$, $\Lambda $, $\rho $,
$a$ and the embedding:
\begin{align}
&\left( \overline{K}+u_{a}u_{b}\overline{K}^{ab}\right) \Biggl[
\frac{ \kappa^2}{\gamma}\left( \rho +3p\right)+2\left( \Lambda
-\kappa ^{2}U\right)
\notag \\
&~{} -\kappa ^{2}{\frac{(1+\gamma)^2}{\gamma^2}}{\lambda }\Biggr]
-3\epsilon\kappa^2{\frac{(1+\gamma)}{\gamma}} \left[ {\frac{
\Delta\widetilde\Lambda }{\widetilde{\kappa}^2}}+
{\frac{\lambda}{\gamma}}\,
\overline{K}\right] \mathcal{A}  \notag \\
&~{}+{\kappa ^{2}}{\frac{(1+\gamma)^2}{\gamma^2}}{\lambda }\left[
-2 \overline{K}+u_{a}u_{b}\overline{K}
^{ab}\right]\mathcal{A}^{2}=0\,. \label{fc1a}
\end{align}
To leading order, this constraint agrees with the corresponding
constraint of the asymmetric RS model~\cite{Decomp}.

The Lanczos equation~(\ref{fLanczos}) becomes
\begin{align}
& \frac{3\epsilon }{\widetilde{\kappa }^{2}}{(1+\gamma
)}\mathcal{A}\,\Delta
K_{ab}=\Bigl[\rho +3p-\frac{(1+\gamma )^{2}}{\gamma }\lambda   \notag \\
& ~{}+{\frac{2\gamma }{\kappa ^{2}}}\left( \Lambda -\kappa
^{2}U\right) -\epsilon {\frac{(1+\gamma )}{\gamma }}\lambda
\mathcal{A}\Bigr]u_{a}u_{b}
\notag \\
& ~{}+{\frac{(1+\gamma )}{\gamma }}\lambda \mathcal{A}\left[
\epsilon +(1+\gamma )\mathcal{A}\right] a^{2}h_{ab}\,.
\label{fLanczosa}
\end{align}

\section{Anti de Sitter bulk}

Since the bulk equations are unchanged by the introduction of the
induced gravity contribution on the brane, the {static} bulk
solution admitting spatial sections with cosmological symmetry
will be AdS. Then the off-brane evolution equations~(\ref{dep}),
(\ref{de}) and (\ref{G2p}) are trivially satisfied and
$\overline{\mathcal{E}}_{ab}=0$.

\subsection{Determining the functions $U$ and $\Lambda $}

From the extrinsic curvature of a generic Friedmann
brane~\cite{Decomp} we find
\begin{equation}
h^{ab}\Delta K_{ab}=6a\overline{B}\,,
\end{equation}
where
\begin{equation}
B^{\pm }=-\left[ \dot{a}^{2}+k-{\frac{\widetilde{\Lambda}^\pm
}{6}}\,a^2 \right] ^{1/2}.  \label{B}
\end{equation}
(We have chosen the normal to the brane pointing towards the $+$
region.) Then
\begin{equation}
12\overline{B}\Delta B+a^{2}\Delta \widetilde{\Lambda }=0\,,
\label{dB}
\end{equation}
and
\begin{align}
&\frac{\dot{a}^{2}+k}{a^{2}}=\frac{\Lambda _{0}}{3}-\frac{\kappa
^{2}\lambda }{6}+\left( \frac{\overline{B}}{a}\right) ^{2}+\left(
\frac{\Delta B}{2a} \right) ^{2}\,,  \label{Friedm}
\end{align}
which is the Friedmann equation, written in terms of
$\overline{B}$ and $ \Delta B$. These can be determined from
Eq.~(\ref{dB}) and the $h^{ab}$ -projection of the Lanczos
equation~(\ref{fLanczosa}):
\begin{eqnarray}
H^2+\frac{k}{a^{2}} &=&\frac{\Lambda _{0}}{3}+\frac{\kappa
^{2}\lambda }{6} \left\{\left[ {\frac{1+\epsilon (1+\gamma)
\mathcal{A} }{\gamma}} \right]
^2-1\right\}  \notag \\
&&~{}+\frac{\gamma^2(\Delta \widetilde{\Lambda })^{2}}{96\kappa
^{2}\lambda[ 1+ \epsilon (1+\gamma)\mathcal{A}]^2}\,.
\label{Friedmann}
\end{eqnarray}
In the $\gamma\rightarrow 0$ limit, using Eq.~(\ref{limit}), we
recover the Friedmann equation of the asymmetric RS model with AdS
bulk~\cite{Decomp}. We have checked that the remaining
constraints, Eqs.~(\ref{fc1a}) and the Codazzi equation, are
satisfied.

A comparison of the generic form of the Friedmann equation,
Eq.~(\ref{Fr}), with the specific form for an AdS bulk,
Eq.~(\ref{Friedmann}), gives an algebraic relation between
$\Lambda $\ and $U$. We obtain a \textit{quartic} polynomial
\begin{equation}
\sum_{i=0}^{4}c_{i}x^{i}=0\,,  \label{pol4}
\end{equation}
in the variable
\begin{equation}
x=\Lambda -\Lambda _{0}+\kappa ^{2}U\,,
\end{equation}
with coefficients
\begin{eqnarray}
c_{0} &=&-{\frac{\gamma^2\lambda^2(\Delta \widetilde{\Lambda
})^{4} }{
4\kappa^4}} \,,  \notag \\
c_{1} &=&{\frac{32\lambda^2 \left( \lambda +\gamma\beta \right)
(\Delta
\widetilde{\Lambda })^{2} }{\kappa^2}}\,,  \notag \\
c_{2} &=&-1024\lambda ^{2}\left[ \frac{\gamma^2(\Delta
\widetilde{\Lambda }
)^{2}}{32\kappa^4 } +\beta ^{2}\right] \,,  \notag \\
c_{3} &=&{\frac{2048\gamma\lambda ^{2}\beta }{\kappa^2}} \,,  \notag \\
c_{4} &=&-{\frac{1024\gamma^2\lambda ^{2}}{\kappa^4}}\,.
\end{eqnarray}
Here we have introduced the notation
\begin{equation}  \label{beta}
\beta =\rho +\lambda\left(1+{\frac{\gamma }{2}}
\right)-{\frac{\gamma\Lambda _{0}}{\kappa^2}}\,.
\end{equation}

Equation~(\ref{pol4}) was obtained by eliminating $\dot{a}^{2}$
from the two forms of the Friedmann equation, Eqs.~(\ref{Fr}) and
(\ref{Friedmann}), then inserting $\mathcal{A}^{2}$ from
Eq.~(\ref{A2}), expressing $\mathcal{A}$ from the resulting
equation, taking its square and comparing again with $ \mathcal{A}
^{2}$ as given by Eq.~(\ref{A2}). With this procedure, the
explicit dependence on $\epsilon $ is lost, and the two branches
will re-emerge as roots of Eq.~(\ref{pol4}).

The functions $U$ and $\Lambda-\Lambda_0 $ obey Eq.~(\ref{U0}),
which becomes
\begin{equation}
\kappa ^{2}U=\frac{3\left( \rho +p\right) \dot{x}}{4\dot{\rho}}\,.
\label{diff}
\end{equation}
Once the solutions of Eq.~(\ref{pol4}) are found, $U $ is given by
Eq.~(\ref {diff}). This is an alternative route to the lengthy
procedure, described in~ \cite{Decomp}, of comparing the two forms
of the Raychaudhuri equation,.

In the $Z_{2}$-symmetric limit, with $\Delta \widetilde{\Lambda
}=0$ ,\ the coefficients of the polynomial~(\ref{pol4}) reduce to
\begin{eqnarray}
c_{0} &=&c_{1}=0\,,~~ c_{2} =-1024\lambda ^{2}\beta ^{2}\,,  \notag \\
c_{3} &=&2048\gamma\kappa^{-2}\lambda ^{2}\beta \,, ~~ c_{4}
=-1024\gamma^2\kappa^{-4}\lambda ^{2}\,.
\end{eqnarray}
The only solution is the double root $x=0$. Then Eq.~(\ref{diff})
is solved for $U=0$ and we recover the result of~\cite{MMT}, in
the AdS case $\mathcal{ E} _{0}=0 $.

\subsection{Perturbative solutions: small asymmetry}

A small deviation from $Z_{2}$-symmetry can be described via the
dimensionless parameter
\begin{equation}
\alpha={\frac{(\Delta \widetilde{\Lambda }
)^{2}}{6\kappa^{4}\lambda ^{2}}} \ll1 \,.
\end{equation}
To leading order, the solution of Eq.~(\ref{pol4}) is
\begin{equation}
x={\frac{3\kappa^2\lambda^2 }{32 \beta^2}} \left[ \lambda
+\gamma\beta -\epsilon \sqrt{\lambda \left( \lambda +2\gamma\beta
\right) }\right]\alpha . \label{xx}
\end{equation}
From Eq.~(\ref{diff}) we obtain
\begin{eqnarray}
\kappa ^{2}U &=&-\frac{9\kappa^2\left( \rho
+p\right)\lambda^{2}}{128 \beta ^{3}\left( \lambda +2\gamma\beta
\right) } \Bigl[\left( \lambda
+2\gamma\beta \right) \left( 2\lambda +\gamma\beta \right)  \notag \\
&&~{}-\epsilon \left( 2\lambda +3\gamma\beta \right) \sqrt{\lambda
\left( \lambda +2\gamma\beta \right) }\Bigr]\alpha.  \label{UU}
\end{eqnarray}

Switching off the induced gravity in this perturbative solution
gives
\begin{eqnarray}
x &=&\frac{3\left( 1-\epsilon \right) \kappa^2\lambda^3}{ 32\left(
\rho
+\lambda \right) ^{2}}\,\alpha\,, \\
\kappa ^{2}U &=&-\frac{9\left( 1-\epsilon \right)\kappa^2 \left(
\rho +p\right)\lambda^3 }{64\left( \rho +\lambda \right)
^{3}}\,\alpha\,.
\end{eqnarray}
The perturbative solution is in fact \emph{exact} in the limit
without induced gravity, as can be verified by direct inspection
of Eqs.~(\ref{pol4} ) and (\ref{diff}) in the $\gamma\rightarrow
0$ limit. For the $\epsilon =-1$ branch the results of the
asymmetric RS model, presented in~\cite{Decomp}, are recovered.

Having obtained $x$ and $U$, the explicit forms of the Friedmann
and Raychaudhuri equations, Eqs.~(\ref{Fr}) and (\ref{Ray}), can
be written. For this, we remark that
\begin{equation}
\mathcal{A}^{2}=1+\frac{2\gamma }{(1+\gamma )^{2}\kappa
^{2}\lambda }\left( \kappa ^{2}\rho -\gamma x-\gamma \Lambda
_{0}\right) .  \label{AA2}
\end{equation}
In the $Z_{2}$-symmetric limit this agrees with Eq.~(3.26)
of~\cite{MMT}, for vanishing dark radiation term.

\subsection{High energy limit}

We now consider the perturbative small-asymmetry case at high
energies ($ \rho \gg \lambda $) in the early universe. We assume
that there is no bare cosmological constant on the brane, i.e.,
$\Lambda _{0}=0$, and that the spatial sections are flat, $k=0$.
By Eqs.~(\ref{beta}) and (\ref{xx}),
\begin{equation}
\beta \approx \rho \,,~~x\approx \left( {\frac{3}{32}}\gamma
\kappa ^{2}\lambda \right) {\frac{\lambda }{\rho }}\,\alpha \,.
\end{equation}
Then from Eq.~(\ref{AA2}),
\begin{equation}
\mathcal{A}^{2}={\frac{2\gamma }{(1+\gamma )^{2}}}\,{\frac{\rho
}{\lambda }} \left[ 1+O\left( {\frac{\lambda }{\rho }}\,,\alpha
{\frac{\lambda ^{2}}{\rho ^{2}}}\right) \right] ,
\end{equation}
so that to lowest order, there is no effect from asymmetry. The
Friedmann equation~(\ref{Fr}) is the same as in the symmetric
case:
\begin{equation}
H^{2}={\frac{\kappa ^{2}\rho }{3\gamma }}\left[ 1+\epsilon
\sqrt{\frac{ 2\lambda }{\gamma \rho }}\left\{ 1+O\left(
\sqrt{\frac{\lambda }{\rho }} \,,\alpha {\frac{\lambda ^{2}}{\rho
^{2}}}\right) \right\} \right] .
\end{equation}

\subsection{Low energy limit}

In the low-energy, late-universe regime, induced gravity models
with $ \epsilon=1$ are interesting because they can produce
acceleration even without dark energy~\cite{d2,ig}. Asymmetric RS
models can also produce acceleration (see~\cite{Decomp} and
references therein). We consider $\epsilon=\pm 1$ models in the
perturbative small-asymmetry case, with $\Lambda_0=0=k$. We assume
that
\begin{equation}  \label{order}
{\frac{\rho}{\lambda}}\ll \alpha \ll 1\,,
\end{equation}
which is readily achieved by choosing the brane tension high
enough relative to the very low energy scale of the late universe.

By Eqs.~(\ref{beta}) and (\ref{xx}),
\begin{eqnarray}
\beta  &\approx &\lambda \left( 1+\frac{\gamma }{2}\right) \,, \\
x &\approx &{\frac{3}{16}}\kappa ^{2}\lambda \left(
{\frac{1-\epsilon +\gamma }{2+\gamma }}\right) ^{2}\alpha \,.
\end{eqnarray}
Then from Eqs.~(\ref{AA2}) and (\ref{order}),
\begin{equation}
\mathcal{A}\approx 1-\frac{3}{16}\left[ {\frac{\gamma \left(
1-\epsilon +\gamma \right) }{(1+\gamma )\left( 2+\gamma \right)
}}\right] ^{2}\,\alpha \,.
\end{equation}
To lowest order, the first correction is from asymmetry, in
contrast to the high-energy case. The Friedmann
equation~(\ref{Fr}) becomes
\begin{equation}
H^{2} ={\frac{2\kappa ^{2}(1+\gamma )\lambda }{3\gamma
^{2}}}\!\left[ 1\!-\!{\frac{3\gamma ^{4}}{32(1+\gamma )^{2}\left(
2+\gamma \right)
^{2}}}\,\alpha \right] ,
\end{equation}
for ${\epsilon =1}$, and
\begin{equation}
H^{2} ={\frac{\kappa ^{2}\lambda }{16(1+\gamma )}} \,\alpha \,,
\end{equation}
for ${\epsilon =-1}$.

It follows that for $\epsilon=1$, the late-time acceleration from
induced gravity is slightly reduced by asymmetry, whereas for the
$\epsilon=-1$ branch (with an RS limit), the asymmetry introduces
a small acceleration.

\section{Concluding remarks}

Combining the method developed in~\cite{Decomp} with that
of~\cite{MMT}, we have developed a general formalism describing
gravitational dynamics on the asymmetrically embedded brane with
induced gravity. The decomposition of the bulk Einstein equations
yielded the tensorial, vectorial and scalar projections. They are
equivalent to the projected Einstein, the Codazzi and the twice
contracted Gauss equations. In the derivation of the final set of
equations the {generalized} junction conditions were applied. The
resulting effective Einstein equation contains all terms from the
asymmetric RS model, to which the induced gravity contributions
add linearly.

For Friedmann branes, the generalized Friedmann and Raychaudhuri
equations were derived, together with the relevant constraint
equations. They contain two undetermined functions functions $U$
and $\Lambda -\Lambda_0$. These depend on the bulk geometry. As an
application of the formalism, we discussed in detail the AdS bulk
with different cosmological constants on the two sides of the
brane. Determining the functions $U$ and $ \Lambda-\Lambda_0$
reduces in this case to solving a quartic polynomial. This feature
of the model was already remarked on and exploited in~\cite{KLM} ,
where a perturbative solution in terms of a small induced gravity
contribution was analyzed. Similarly, in~\cite{Shtanov} a quartic
equation in $H^{2}+k/a^{2}$ was presented. Our method in
determining the Friedmann equation also yielded a quartic
equation. We have found and discussed those solutions of the
quartic equation for which the deviation from $Z_{2}$ -symmetry is
small. In the high-energy, early-universe regime, the asymmetry
does not affect the Friedmann equation to lowest order. In the
low-energy, late-universe regime, the asymmetry does have an
effect. In the $\epsilon=-1$ branch, it contributes a late-time
acceleration, while in the $\epsilon=1$ branch is slightly reduces
the late-time acceleration arising from induced gravity.

\newpage
{\bf Acknowledgments}

We thank Cedric Deffayet for useful comments. L\'{A}G was
supported by the Hungarian Scholarship Board, OTKA grants Nos.
T046939 and TS044665 and a PPARC visitor grant at Portsmouth. RM
is supported by PPARC.

\end{document}